\documentclass[aps,pre,twocolumn,amsmath,amssymb,nofootinbib,floatfix,superscriptaddress]{revtex4-1}

\usepackage{amsmath,braket}
\usepackage{amssymb}
\usepackage{amsthm}
\usepackage[dvipdf]{color}
\usepackage{graphicx}
\usepackage{dcolumn}
\usepackage{bm,subfigure} 
\usepackage{xcolor,colortbl}
\usepackage[normalem]{ulem}
\usepackage{circuitikz}

\usepackage{enumerate}
\usepackage{enumitem}

\begin{document}

\title{Statistical mechanics of frustrated assemblies and incompatible graphs}

  \author{José M. Ortiz-Tavárez}
  \thanks{These  authors contributed equally}
   \affiliation{
 Department of Physics,
  University of Michigan, Ann Arbor, 
 MI 48109-1040, USA
 }
  \author{Zhen Yang}
  \thanks{These  authors contributed equally}
   \affiliation{
 Department of Physics,
  University of Michigan, Ann Arbor, 
 MI 48109-1040, USA
 }
   \author{Nicholas Kotov}
 \affiliation{
 Department of Chemical Engineering, University of Michigan, Ann Arbor,
USA
 }
 \affiliation{Department of Materials Science, University of Michigan, Ann Arbor, USA}
 \affiliation{Center for Complex Particle Systems (COMPASS), University of Michigan, Ann Arbor, USA}
  \author{Xiaoming Mao}
 \affiliation{
 Department of Physics,
  University of Michigan, Ann Arbor, 
 MI 48109-1040, USA
 }
  \affiliation{Center for Complex Particle Systems (COMPASS), University of Michigan, Ann Arbor, USA}

\begin{abstract} 
Geometrically frustrated assemblies where building blocks misfit have been shown to generate intriguing phenomena from self-limited growth, fiber formation, to structural complexity. We introduce a graph theory formulation of geometrically frustrated assemblies, capturing frustrated interactions through the concept of incompatible flows, providing a direct link between  structural connectivity and frustration.  This theory offers a minimal yet comprehensive framework for the fundamental statistical mechanics of frustrated assemblies.  Through numerical simulations, the theory reveals new characteristics of frustrated assemblies, including two distinct percolation transitions for structure and stress, a crossover between cumulative and non-cumulative frustration controlled by disorder, 
and a divergent length scale in their response. 
\end{abstract}
\maketitle

\noindent{\it Introduction.}---Complex structured materials, such as seashell nacre, articular cartilage, bones, and tree roots, combine order and disorder and can offer functionalities that surpass perfectly ordered and completely disordered materials~\cite{blank2003nacre,gilbert2022biomineralization,durkovic2020cell,jiang2020emergence,mao2024complexity}.  These materials often exhibit complex features such as layers, fibers, and clusters that intertwine and evolve across multiple length scales.   
Geometrically frustrated assembly (GFA)~\cite{bruss2012non,grason2016perspective,haddad2019twist,li2020some,meiri2021cumulative,sadoc2020liquid,PhysRevE.104.054601,serafin2021frustrated,schonhofer2023rationalizing}, where building blocks misfit, has been proposed as a mechanism that generates structural complexity with minimal energy expenditures~\cite{serafin2021frustrated,mao2024complexity}.  

Geometric frustration has been a fundamental concept in physics and has led to a wealth of complex phenomena from glassy dynamics~\cite{binder1986spin} to topological order~\cite{Broholm2020}.  In the context of self assembly, it is well known that geometric frustration can lead to self-limiting clusters~\cite{grason2016perspective,meiri2021cumulative} and topological defects~\cite{grason2012defects}, based on minimizing energies.  The statistical mechanics of GFAs has just started to be studied, revealing complex phase diagrams resulting from the cooperation of entropy and frustration energy~\cite{hagan2021equilibrium,tyukodi2022thermodynamic,PhysRevX.13.041010}.

In this paper, we introduce a graph theory formulation for the statistical mechanics of GFAs, where mechanical stress due to geometric frustration maps to incompatible flows on graphs, not only enabling automatic minimization of the elastic energy but also elucidating the relation between stress and the connectivity of the assembly.  
The theory reveals curious features including the generation of branched structures, two percolation transitions for structure and stress, disorder induced crossover, 
and long-ranged cooperation resembling self-organized criticality~\cite{bak1988self}, shedding light on the fundamental structural complexity of GFAs.

In particular, we investigate models of different types of geometric frustration: assemblies with non-cumulative (characterized by an extensive frustration energy) and cumulative (characterized by a super-extensive frustration energy) frustration, following the classification introduced in Ref.~\cite{meiri2021cumulative}, and the combination of the two. Interestingly, disorder serves as a thermodynamic control parameter for the crossover between the two cases.  

\begin{figure}[h]
    \centering
    \includegraphics[width=0.4\textwidth]{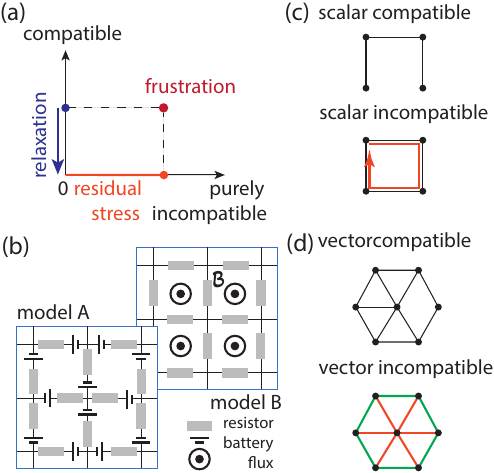}
    \caption{(a) Decomposition of frustration to the two linear subspaces in the edge space: the compatible where the potential drop on the edges can be attributed to node potential differences (the edge rest length can be satisfied by node displacements) and the incompatible where they can not for the scalar (vector) version. (b) Illustration of Models A and B.  (c-d) Examples of graphs without and with incompatible flows for the scalar (c) and vector (d) problems.  }
    \label{fig:incompatible}
\end{figure}

\noindent{\it The analogy between frustrated elasticity and incompatible flows.}---
Consider a set of building blocks, for which the ideal metric of local fitting is represented by  $\bar{g}$, which is in general non-Euclidean and encodes the information about the frustration.
This material is then ``flattened'' to the Euclidean space represented by metric $I$ (identity).  This flattened state is an arbitrarily chosen reference state (with the only requirement being Euclidean) and in general does not minimize the elastic energy.  The energy-minimizing state, being also Euclidean, is related to this reference state by a displacement field $\vec{u}$.  The total strain of the energy-minimizing state from the non-Euclidean ideal state,
\begin{equation}\label{eq:strain}
    \epsilon_{ij} = \frac{1}{2}\left(
    \partial_i u_j +\partial_j u_i
    \right) + (I-\bar{g})_{ij} ,
\end{equation}
contains a compatible component (first term) and an incompatible component (second term) which can not be written in terms of $\vec{u}$ if $\bar{g}$ is non-Euclidean (Fig.~\ref{fig:incompatible}).  In two dimensions this is straightforward to see, as the the last term is related to the gaussian curvature~\cite{Ciarlet2008}, $\varepsilon_{ik}\varepsilon_{jk}\partial_k \partial_l \bar{g}_{ij}=2K$ where $\varepsilon$ is the Levi-Civita symbol.  This is also similar to the strain field of disclinations in crystals where curvature focuses~\cite{chaikin1995principles}.  

We utilize the analogous structure of Eq.~\eqref{eq:strain} with the equation for electric fields in terms of the electromagnetic potential $(\phi,\mathcal{A})$,
\begin{equation}\label{eq:efield}
    \mathcal{E}_i = -\partial_i \phi -\dot{\mathcal{A}}_i .
\end{equation}
Similarly, $\nabla\times\mathcal{A}=-\mathcal{B}$ (the magnetic field) can not be written in terms of $\partial_i \phi$.  This analogy provides a simple scalar version of the vector frustrated elasticity problem. 
Similar analogies have been utilized for emergent elasticity in amorphous solids~\cite{nampoothiri2020emergent,PhysRevE.106.065004} and fracton-elasticity duality~\cite{PhysRevLett.120.195301}.  
We study Eq.~\eqref{eq:efield} on graphs as electric circuits, and show that the existence of incompatible flows is characterized by the linear space of cycles, in the same way where incompatible strains are characterized by the linear space of states of self stress (SSSs)~\cite{sun2012surface,mao2018maxwell,sun2020continuum,sun2019maxwell,PhysRevLett.127.098001,zhang2022prestressed}, allowing convenient statistical mechanics modeling of complex structured GFAs. 



\begin{figure*}[t]
    \centering
    \includegraphics[width=0.95\textwidth]{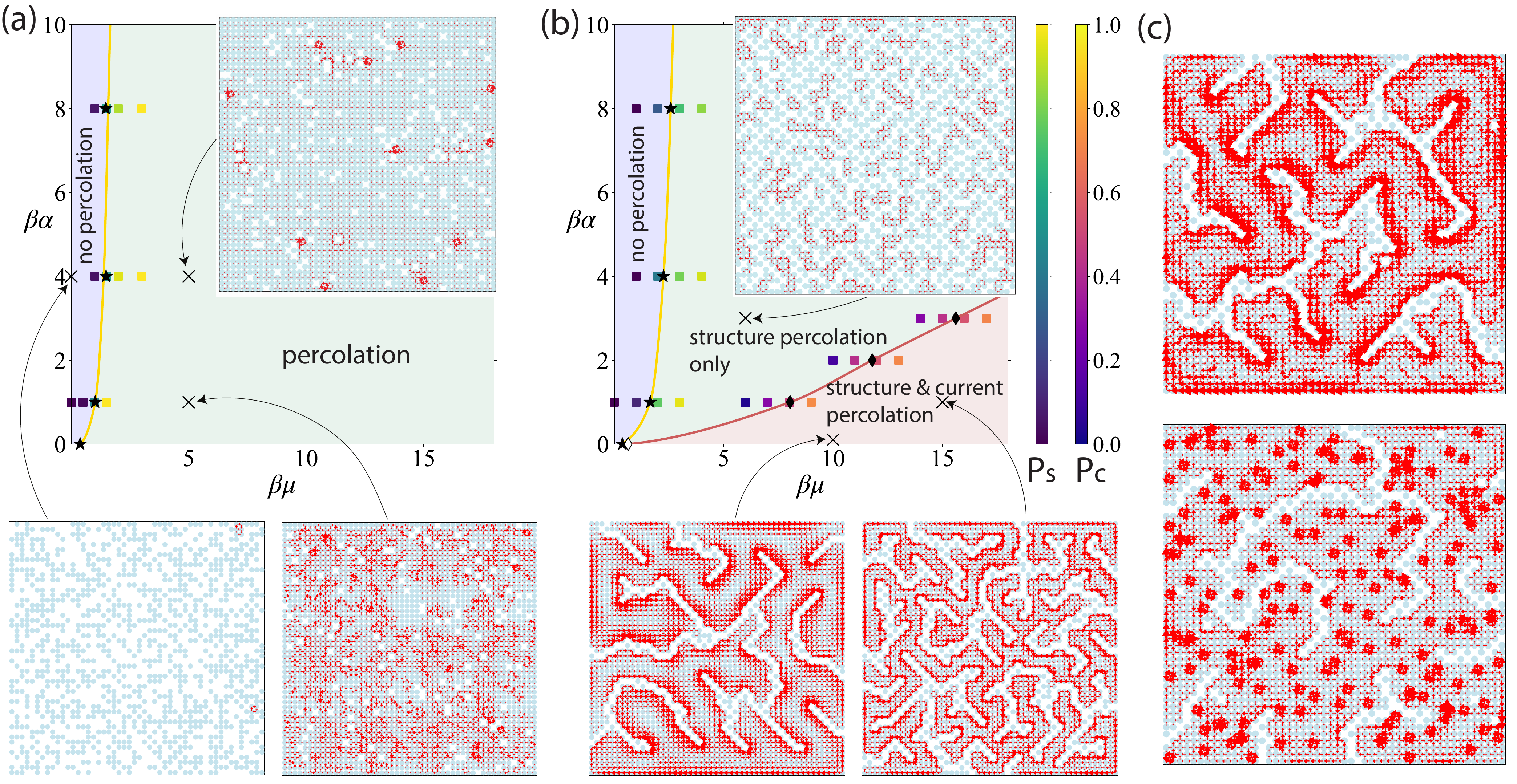}
    \caption{(a-b) Phase diagrams of Model A (a) and Model B (b), with phase boundaries (stars and diamonds) determined from the probability of structure ($P_s$) and current ($P_c$) percolation (data points with color bars in (b), as detailed in the SM).  Three examples of configurations ($(\beta\mu,\beta\alpha)=(5,4),(0,4),(5,1)$ for (a) and $(\beta\mu,\beta\alpha)=(6,3),(10,0.1),(15,1)$ for (b)) are shown with arrows pointing to the parameters (X's) on each phase diagram.  Particles and currents are shown as gray disks and red arrows  with the thickness of the arrows proportional to the amplitudes of the current normalized by $\beta\alpha$.  (c) Examples of Model C configurations at $(\beta\mu,\beta\alpha)=(10,0.1)$ and $\gamma=1$ (top) and $\gamma=6$ (bottom). }
    \label{fig:phasediagrams}
\end{figure*}

\noindent{\it Incompatible flow graphs.}---Consider a graph $G$ with $N_{\textrm{nodes}}$ nodes and $N_{\textrm{edges}}$ edges.  Let $C$ be its $N_{\textrm{edges}} \times N_{\textrm{nodes}} $ dimensional incidence matrix~\cite{strang2016introduction}.  
The null space of $C^T$ defines a $N_\textrm{cycles}$ dimensional linear space of edge flows $\mathbf{i}$ without 
sources or sinks at any nodes, i.e., cycle flows.  
Choosing a set of vectors in this null space such that they form a basis for it and assembling them in to a matrix, we obtain a $N_\textrm{cycles} \times N_{\textrm{edges}}$ dimensional matrix $B$, the circuit matrix (as in \cite{swamy1981graphs,SHAI2001343,Lloyd1978GraphTW}). 
Note that a direction convention needs to be chosen in constructing these matrices so that $\pm$ sign of the edge flows are defined relative to their reference direction.

The circuit matrix $B$ contains information that is essential to the discussion of geometric frustration.  Its row space and null space define two orthogonal linear subspaces of the $N_{\textrm{edges}}$ dimensional edge space: the $N_\textrm{cycles}$ dimensional ``purely incompatible'' and the $N_\textrm{nodes}-N_{\textrm{c.c.}}$ dimensional ``compatible'' flows, respectively.  The meaning of the names can be understood by considering imposed potential drops $\mathbf{w}$ on the edges, e.g., installing batteries on the graph as an electric circuit and assuming a resistance $r$ on each edge in series with the battery. The physical potential drop on each edge is $\Delta \mathbf{v} = C \mathbf{v} = \mathbf{w}-r\mathbf{i}$, where $\mathbf{v}$ are the node potentials.  Using orthonormalized $B$, as detailed in the Supplementary Information (SM), we can decompose the edge space into the compatible and the purely incompatible components, 
\begin{equation}\label{eq:ri}
   r \mathbf{i} = - C\mathbf{v} + \mathbf{w} 
   = \left( -C\mathbf{v} +(1-B^T B) \mathbf{w} \right) +  B^T B \mathbf{w} .
\end{equation}
In a circuit without sources or sinks, $\mathbf{i}$ can only contain cycle currents, $C\mathbf{v} +(1-B^T B) \mathbf{w}=0$, meaning that the system reacts with node potentials $\mathbf{v}$ to relax $\mathbf{w}$.  The term that can not be relaxed away, $B^T B \mathbf{w}$, is called incompatible.  As a result, the power dissipation of the system is 
\begin{equation}\label{eq:power}
    P = r \mathbf{i}^T\mathbf{i} =
    \frac{1}{r} \mathbf{w}^T B^T B\mathbf{w}
    = \frac{1}{r} \vert B \mathbf{w} \vert ^2 .
\end{equation}

It is interesting to note that the $r \mathbf{i} = - C\mathbf{v} + \mathbf{w} $ part in Eq.~\eqref{eq:ri} is the graph analog of Eq.~\eqref{eq:strain} for elasticity and Eq.~\eqref{eq:efield} for the electric field.  In all cases, the ``frustration'' ($\mathbf{w}, I-\bar{g}, \mathcal{A}$) can be decomposed into the compatible part which can be relaxed away through the first term (by the choice of $\mathbf{v},u,\phi$ respectively), and an incompatible part which can't.  The graph theory provides convenient tools for this decomposition.

The notion of incompatibility is even clearer in the vector version of this problem, which corresponds to stress in mechanical frames.  
This readily follows by simply generalizing the $C,C^T$ matrices to the mechanical compatibility and equilibrium matrices~\cite{pellegrino1986matrix,sun2012surface}, where the dimensions $N_\textrm{nodes}, N_{\textrm{cycles}}, N_{\textrm{c.c.}}$ turn into the numbers of degrees of freedom, SSSs and zero modes $N_\textrm{nodes} d, N_{SSS}, N_{0}$ where $d$ is the spatial dimension, whereas $N_{edges}$ stay the same.  
The circuit matrix $B$ is then generalized to a $N_{SSS}\times N_{\textrm{edges}}$ matrix containing all the SSSs $\mathbf{t}$ (defined as $C^T \mathbf{t}=0$) as row vectors.  The meaning of ``compatibility'' is evident in this case.  Starting from a stress-free configuration where all edges are at their rest lengths, we assign changes to the rest length (``batteries'') $\mathbf{w}$ to the edges. 
The compatible part of the rest length changes, $(I-B^T B)\mathbf{w}$, can be relaxed by nodes adopting new equilibrium positions, but the incompatible part $B^T B\mathbf{w}$ belongs to the SSS (``cycles'') space and can not be relaxed,  
resulting in nonzero prestress in equilibrium.  In a physical self-assembly system, the rest lengths are defined by the interaction potentials between the building blocks: the system runs into geometric frustration only when the minimum interaction potentials pair distances projects to the SSSs space (Fig.~\ref{fig:incompatible}d).  

It is worth noting that due to the convenient choice of dynamical equations on the graphs, the incompatibility $|B\mathbf{w}|^2$ corresponds to dissipation in the scalar model but stored elastic energy in the vector model, but the essential step of decomposing driving forces on the edges is common and can be extended to more general cases.




\noindent{\it Models.}---Based on the formulation developed above, we propose three models of GFAs on square lattices, and apply Monte Carlo (MC) simulations at fixed chemical potential $\mu$ and temperature $T$  to these models.  We consider an $L\times L$ square lattice with open boundary conditions, with the total energy being the sum of the power (Eq.~\eqref{eq:power}) and $-\mu N$ from the fixed chemical potential $\mu$ ensemble, which can be written as
\begin{equation}
    E= \alpha \vert B\mathbf{w} \vert^2 -\mu N,
\end{equation}
where $N$ is the particle number (occupied nodes) with edges existing only when both nodes they connect to exist (similar to site percolation models), and run MC simulation using the Metropolis algorithm, as detailed in the SM.  
We discuss these models in the context of the scalar problem, but this formulation is readily generalizable to vector and tensor problems at higher dimensions for GFAs as we discussed above. 

Model A assumes random batteries on the edges, representing random ``misfits'' of the building blocks (Fig.~\ref{fig:incompatible}b).  This corresponds to the case of non-cumulative frustration defined in Ref.~\cite{PhysRevE.104.054601,PhysRevE.105.024703}.
At each MC step, a trial move of adding or removing a node is generated, where each edge $e$ connecting the new node with an existing node is randomly assigned with batteries  $w_e=\pm 1$ with equal probabilities.  The frustration in this model is similar to well-understood cases of spin glasses, where the frustration energy is extensive~\cite{binder1986spin}.

Model B assumes a ``coherent'' frustration, representing systematic misfits caused by non-Euclidean geometries of the building blocks.  This corresponds to the case of cumulative frustration, and is represented by a constant time variation of the $\mathcal{\vec{B}}$ field as we discussed above.  This model shares  similarities with  Ref.~\cite{PhysRevX.13.041010} but with important differences.  Instead of starting from a superconductor hamiltonian and taking the potential on the nodes as thermodynamic variables, we assume a classical gauge field model, where the system is considered as a normal conductor, and a homogeneous magnetic field $\mathcal{\vec{B}}$ of constant time variation is imposed perpendicular to the lattice.  This leads to coherent electromotive forces on the edges, $w_e =- \dot{\vec{\mathcal{A}}} \cdot \vec{\ell}_e$ where $\vec{\mathcal{A}}$ is the vector potential and $\vec{\ell}_e$ is the vector pointing from the initial to the final node of edge $e$.
We take the Landau gauge in which $\vec{\mathcal{A}}=\mathcal{B}_z x \hat{y}$
where $x,y$ are the cartesian coordinates of the edge, $\hat{y}$ is the unit vector along $y$, and $\mathcal{B}_z=1$ is taken so the frustration is controlled by $\alpha$.  This gauge field causes  currents on all cycles, and the amplitude of the flow is proportional to the area of the cycle, penalizing cycles of large areas, causing a frustration energy superextensive to the size of the clusters.  A filled square of side length $L_s$ has frustration energy $E_f\propto L^4_s$.  
As a result, this favors branched assemblies in a thermodynamic setting.


Model C is a combination of A and B, where a gauge field exists at the same time with random batteries, $w_e = -\dot{\vec{\mathcal{A}}} \cdot \vec{\ell}_e \pm \gamma$ where $\gamma$ controls the relative strength of the random frustration to the gauge field and $(+,-)$ are taken equally likely.   This represent the case where the building blocks have both cumulative frustration that causes gaussian curvature, and also random imperfections such as polydispersity.

It is worth noting that the square lattice is a simple choice for the graph model and not a literal representation of the assembly geometry.  Non-Euclidean lattices, such as $\{5,3\},\{7,3\},\{4,5\}$ tilings (in terms of Schläfli  symbols), are represented by gauge fields (Model B and C) on the lattice.

\noindent{\it Results.}---In general, we found that geometric frustration shifts the geometric percolation transition to higher chemical potentials (Fig 2). However, the morphologies and dynamics of the different models show different features.  We use two dimensionless combinations of the parameters $\beta \mu, \beta \alpha$, the temperature normalized chemical potential and frustration strength, for the following discussions.

In Model A, the frustration energy increasingly focuses  only at small cycles in the bulk at large $\beta \alpha$, indicating that the system is able to eliminate to a large extent the frustration by placing random $\mathbf{w}$ in compatible ways (e.g., mix-and-match polydisperse building blocks).

In Model B, such route to eliminate frustration is no longer available.  Due to the cumulative frustration from the gauge field, any cycle causes frustration, and the system is forced to grow complex branching structures to avoid large area cycles.  
Intriguingly,  beyond  geometric percolation,  there is a region in the phase diagram where the structure percolates but the current doesn't, as the system accommodates more particles by growing trees.  This corresponds to growing floppy to isostatic structures without SSSs in the elasticity problem.  
This lasts until large $\beta\mu$ where current/stress has to percolate.

Model C represents an interesting crossover between A and B, where increasing randomness $\gamma$ gradually shifts from  edge flow to bulk defect flow. 
This indicates that disorder can alleviate cumulative frustration in assemblies by counter balancing the stress (SM).

\begin{figure}[h]
    \centering
    \includegraphics[width=0.49\textwidth]{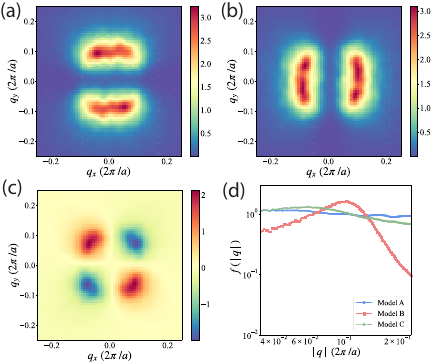}
    \caption{(a-c) Current-current correlation functions $\langle i_x(-q) i_x(q) \rangle$ (a), $\langle i_y(-q) i_y(q) \rangle$ (b), $\langle i_x(-q) i_y(q) \rangle$ (c). (d) The radial part $f(|q|)$.   The data is taken at $(\beta\mu,\beta\alpha)=(10,0.1)$.}
    \label{fig:stress}
\end{figure}

Stress correlation functions have been shown to reveal key structures in prestressed systems~\cite{nampoothiri2020emergent,PhysRevE.106.065004,vinutha2023stress}, so
we examine correlation functions of the currents $\langle \vec{i}(r) \vec{i}(r') \rangle$ in these models (Fig.~\ref{fig:stress}).  The angular dependence of the flows are constrained by the no source/sink rule of the models, $\partial_j i_j(r)=0$, so $\langle \vec{i}(-q) \vec{i}(q) \rangle = (I-\hat{q}\hat{q}) f(|q|)$.  The  models exhibit different features in the radial part $f(|q|)$, demonstrating homogeneity of GFAs generated from non-cumulative frustration and characteristic length scales of stress correlations with cumulative frustration.

Moreover, given the strong cooperatively of the assembly in Model B in avoiding large cycles, we design a test of response length scales by inserting one particle at a random void on the boundary, mimicking new particle attachment in the assembly, and calculate the increase in the energy and the response.  In Model A, similar to other disordered lattice models, such increase is of $\mathcal{O}(1)$.  In Model B, however, this leads to a response at the level of system size: the immediate increase of energy $\Delta E \sim L^{1.7}$, and subsequent dynamics in   minimizing $E$ causes rearrangements at an apparently divergent length scale.  
Because this divergent response occurs over a large area on the phase diagram instead of close to phase transitions, this resembles the response of systems exhibiting self-organized criticality~\cite{bak1988self}.  Importantly, the divergent response is caused by frustration instead of non-equilibrium kinetic rules in this case.  

\begin{figure}[h]
    \centering
    \includegraphics[width=0.49\textwidth]{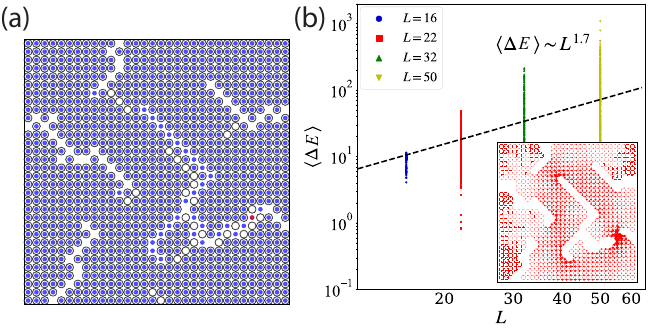}
    \caption{(a) A new particle (red) is added and held at a random void to an equilibrium configuration (blue disks) in Model B.  Energy minimization at $T=0$ of the system leads to long range rearrangement events, arriving at the new equilibrium configuration (black circles). (b) The immediate increase of total energy as a result of adding a particle at a random void at the boundary.
    }
    \label{fig:adding}
\end{figure}





\noindent{\it Acknowledgements.}---The authors thank Siddhartha Sarkar, Nan Cheng,  Mark Newman, Bulbul Chakraborty, and Kai Sun for helpful discussions.  
This work was supported in part by 
the National Science Foundation Center for Complex Particle Systems~(Award \#2243104) and National Science Foundation Award \#2026825, and 
the Office of Naval Research (MURI N00014-20-1-2479).





\appendix

\section{Graph theory for geometrically frustrated assemblies}
\subsection{Potentials and flows on graphs}
    The mathematical framework underlying our work here is the theory of potential graphs and flow graphs. Let us begin with some definitions. A \textbf{graph} is a set nodes and edges that connect them. A \textbf{potential} on a graph is a numerical quantity on the nodes of the graph. A \textbf{flow} on a graph is a numerical quantity assigned on the edges such that $f_{ij}=-f_{ji}$. A \textbf{potential difference} on edge $(i,j)$ refers to the difference between the potential of node $j$ and node $i$.
    These quantities, potentials and flows can be scalars, vectors or even tensors. We will focus here on the scalar problem first.

    In general, we can separate the space of all flows into two components. In a physics context, such as electric circuits, it is useful to describe these two components as flows that satisfy Kirchhoff's cycle rule, and flows that satisfy Kirchhoff's node rule. In graph terms, these two components are referred to as the cut space and cycle space, respectively.
    Any flow that satisfies the cycle rule,
    \begin{equation}\label{eq:cycle}
        0 = \sum_{ij \in \text{cycle}} f_{ij} \quad \forall \text{cycles} ,
    \end{equation}
    is a potential difference and vice versa. Potential differences $\mathbf{\Delta v}$ are obtained from potentials $\mathbf{v}$ via the incidence matrix $C$,
    \begin{equation}
        \mathbf{\Delta v}=C \mathbf{v}.
    \end{equation}
    We refer to these flows as \textbf{compatible} since they  can be realized as potential diferences. This can be compared with the definition of compatible strains in continuum elasticity where a compatible strain tensor comes from derivatives of a displacement field  as we discuss in the main text.\\
    Any flow that satisfies the node rule  Eq.~\eqref{eq:node_rule}, which is to say there is no net flow into any node, can be represented as a linear combination of cycles. In fact, this space is precisely the cycle space of the graph. When a flow has a non-zero projection on the cycle space, we call it \textbf{incompatible}
    \begin{equation}
        0 = \sum_j f_{ij} \quad \forall i. \label{eq:node_rule}
    \end{equation}
       
    Now we introduce the \textbf{circuit matrix} $B$ which we can define as a matrix such that its rows form a cycle basis. By definition the cycle space, balanced flows, is the row space of $B$ and the cut space, compatible flows, is its null space. Note that the two spaces are \textbf{orthogonal complements}. Now the cycle rule (Eq.~\eqref{eq:cycle}) can be written as   
    \begin{equation}
        \mathbf{0}=B\mathbf{f}.
    \end{equation}    
    If we choose the rows of $B$ as an ortho-normal cycle basis then $B^T B$ becomes a projection in to $null(C^T)$
    
    \begin{figure}
        \centering
        \includegraphics[width=0.4\textwidth]{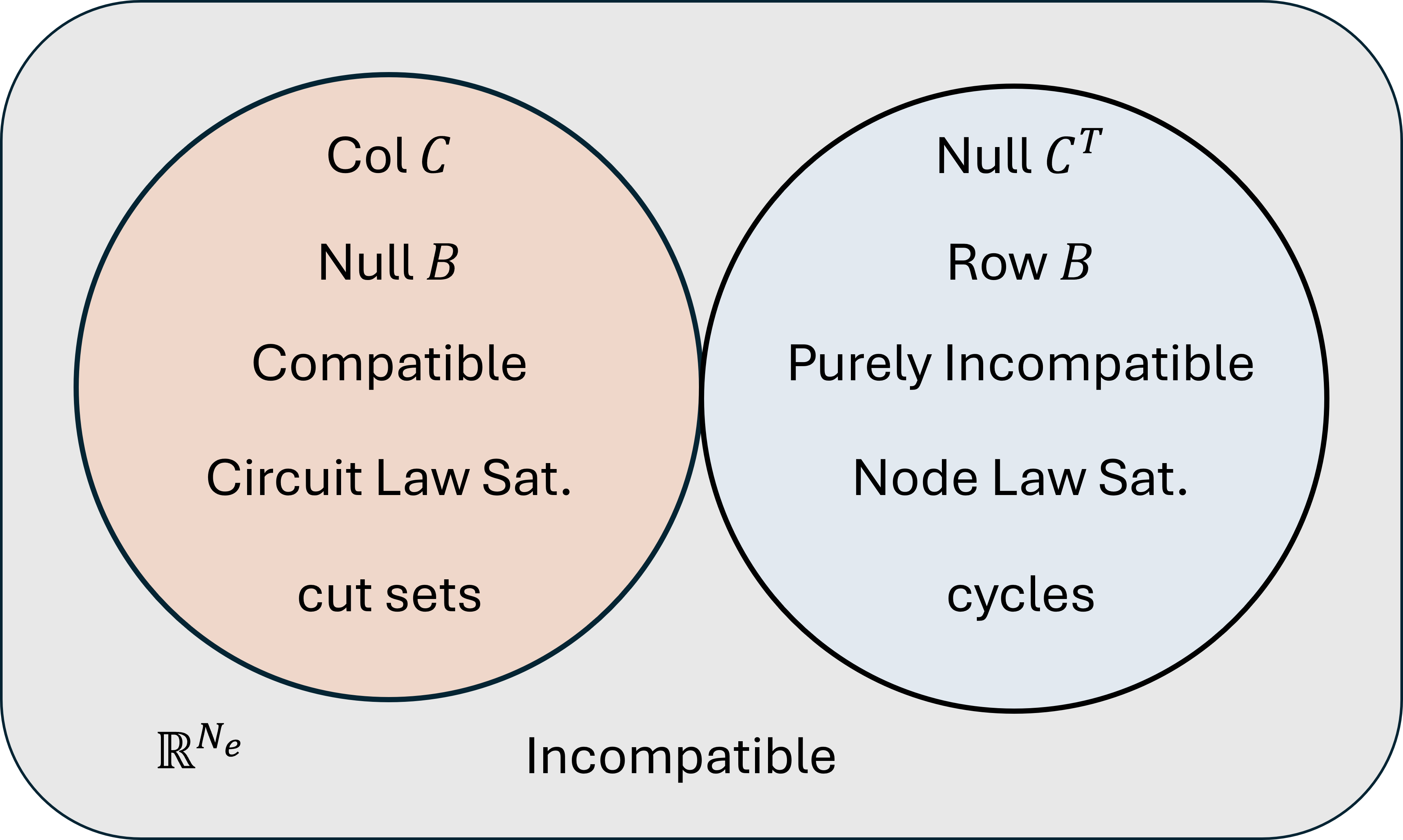}
        \caption{A venn diagram illustrating the relationship between two linear subspaces. The texts inside the circles are all equivalent descriptions of the space. The spaces represented by the two circles are orthogonal complements. They are tangent indicating they share only one element, the zero vector. The space outside both circles represents linear combinations of the two with a nonzero projection on both.}
        \label{fig:venn}
    \end{figure}

    \subsection{Model A: Battery networks}

    \begin{center}
        \begin{circuitikz}[american voltages]
        \draw
          (0,0) to [battery1] (2,0)
          to [R] (4,0);
        \end{circuitikz}
    \end{center}

    Consider an electrical network made up of edges which are a battery with voltage $w_{i,j}$ and internal resistance $r$ (one such edge is shown above). By Ohm's law, 
    the power consumed by the network is given by
    \begin{equation}
        P=r \sum_{ij \in Edges} i_{ij}^2 .
        \label{ohms}
    \end{equation}
    Currents must satisfy the current law,
    \begin{equation}
        C^T \mathbf{i} =0.
    \end{equation}

    Note that if the voltages $\mathbf{w}$ were elements of $col(C)$, they would be compatible, meaning that there would exist a set of potentials at the nodes such that all its differences correspond to the battery voltages which would imply 0 currents for all edges. 
    If this is not the case, i.e., if $\mathbf{w}$ contains components in $null(C^T)$, 
    currents will arise.  
    In particular, the currents arise in such a way as to ``remove'' the incompatibility by creating potential drops across the resistors, so that the final total voltages on the edges are indeed potential differences. In other words the currents must fulfil the following  condition
    \begin{equation}
        0=B^T B\left(-r\mathbf{i} + \mathbf{w}\right).
    \end{equation}

    Note that $B^T B$ projects on to $null(C^T)$, since the currents satisfy the node law they fall completely within this space and so this operator acts on $\mathbf{i}$ as the identity. From the above equation it is then clear $\mathbf{i}r=B^T B \mathbf{w}$. Then by substitution in eq.\ref{ohms} and remembering that $BB^T$ is the identity we obtain 
    \begin{equation}
        P=\frac{1}{r} \mathbf{w}^{T} B^{T} B \mathbf{w} . \label{powerfinal}
    \end{equation}
which is Eq.~(4) in the main text.

\subsection{Model B: Networks with a gauge field}

    Given a planar electric circuit one can consider the effect of a changing magnetic field in the direction orthogonal to the plane. Any closed loop in the circuit will have an induced emf due to the changing magnetic flux. The electric field and the vector potential should satisfy

    \begin{equation}
        \oint (\mathcal{E}+\partial_t \mathcal{A}) \cdot d \boldsymbol{\ell} = 0 ,\label{eq:cont_vectorp_loop}
    \end{equation}
    where the vector potential is related to the magnetic flux enclosed via
    \begin{equation}
        \oint \mathcal{A} \cdot d  \boldsymbol{\ell} = \Phi_{enc}. \label{eq:cont_flux}
    \end{equation}

    To discretize the problem we represent the induced emf on an edge as $a=-\partial_t \Vec{A} \cdot d \Vec{l}$. This emfs play the same role as the batteries played in our previous discussion. The discrete version of Eq.~\eqref{eq:cont_vectorp_loop} is 
    \begin{equation}
        B\left( -r\mathbf{i} + \mathbf{a} \right) = 0.
    \end{equation}
    Given that all edges have the same resistance $r$ we have
    \begin{equation}
        \mathbf{i}=\frac{1}{r}\hat{B}^T\hat{B}\mathbf{a}.
    \end{equation}
    The discrete version of Eq.~\eqref{eq:cont_flux} is
    \begin{equation}
        \boldsymbol{\phi}=B\mathbf{a}.
    \end{equation}
    Where the vector $\boldsymbol{\phi}$ is related to the time derivative of the flux but it depends on the choice of basis for $B$. If a basis is chosen such that each row is a face with appropriate sign then each element $\phi_i$ is the rate of change of the flux on face $i$.  Nevertheless, the resulting current and power on the network is physical and uniquely determined.

    \subsection{Mechanical networks and the vector problem}

        One might consider flows and potentials on a graph which are vectors and define compatible flows analogously to the scalar case. To study the mechanics of spring networks one might choose the position of the nodes as a potential and then tensions on the edges would be an example of a flow. This however is not completely analogous to the scalar version of the problem, electric circuits for example. One significant difference must be noted when applying this formulation to mechanical spring-mass networks: 
        The constitutive relation that relates flows to potential differences is Hooke's law which is not linear with respect to the potential difference $\Delta\mathbf{x}_{ij}= \mathbf{x}_j-\mathbf{x}_i$ where $\mathbf{x}_i,\mathbf{x}_j$ are the positions of nodes $i,j$. Hooke's law can be written as
        \begin{equation}
            \mathbf{t}_{ij}=-k_{ij}\left( \Delta \mathbf{x}_{ij} - l^0_{ij}\frac{\Delta\mathbf{x}_{ij}}{|\Delta\mathbf{x}_{ij}|}\right)
        \end{equation}
        where $l^0_{ij}$ is the rest length of edge $ij$. Note that here we are writing the tension on an edge as a vector $\mathbf{t}_{ij}$ since we are describing it as a ``vector flow" and the position $\mathbf{x}_i$ as the ``vector potential" (not to be confused with the concept of vector potential in E\&M). Note that $\Delta\mathbf{x}_{ij}/|\Delta\mathbf{x}_{ij}|$ reduces to $\pm 1$ for one dimensional (1D) networks making the relationship linear. This 1D mechanical spring network problem is in fact exactly analogous to the battery network problem we described earlier. The battery voltages are the natural length of the springs, the resistance is one over spring constant. Electrical potential becomes position is space, currents become tensions. The voltage drop across the resistors will be analogous to the elongation of the springs from their natural length. 
        
        For dimensions $d\ge 2$  Hooke's Law is linear only to first order on the deformation from some initial configuration. And so all the results of the 1D case are generalizable to arbitrary dimensions but only as first order approximations. In the context of small deviations from equilibrium one can take node displacements as a potential and elongations and tensions as the flows. Since the displacements are assumed small the direction of the edges, their geometry, is assumed to not change appreciably. It is then convenient to package the geometric information with the incidence matrix in to a new matrix called the compatibility matrix.
         For the mechanics of discrete frames we can define matrices $C$ and $Q=C^T$, compatibility and equilibrium matrices respectively~\cite{mao2018maxwell}. The compatibility matrix has as many rows as there are edges in the network and one column per node degree of freedom ($d N$ for $N$ nodes in $d$ dimensions). When $C$ acts on a set of infinitesimal displacements we obtain $\mathbf{e}=C\mathbf{u}$ a set of edge elongations. We can also define a matrix $B$ whose rows form a basis for $null(C^T)$. For spring networks in 1D the compatibility matrix is just the incidence matrix. In two or more dimensions $C$ depends on the geometry of the edges meaning their orientation in space. The role of all these matrices is then analogous to their 1D or scalar problem counterparts.\\
          Let $\mathbf{w}$ be a vector of small changes in rest length (or rates of change) and $\mathbf{u}$ the resulting small displacements (or velocities). Then
          \begin{equation}
            -K(C\mathbf{u}-\mathbf{w})=   \mathbf{t},
        \end{equation}    
        where $K$ is a diagonal matrix whose elements are the spring constants. The total elongation comes from adding the effects of changing rest-lengths and the displacement of the nodes and so $\mathbf{e}=C\mathbf{u}-\mathbf{w}$. Since these are Hookean springs the total energy is given by        
        \begin{equation}
            E=-\frac{1}{2} \mathbf{e}^T  \mathbf{t} .
        \end{equation}    
        By substituting $\mathbf{t}= B^TB\mathbf{t}$ and $\mathbf{e}=C\mathbf{u}-\mathbf{w}$  we will arrive at equation    
        \begin{equation}
            E=\frac{1}{2} \mathbf{w} B^T  \left(  B K^{-1} B^T \right)^{-1}  B\mathbf{w}^T.
        \end{equation}    
        If all spring constants are equal this reduces to    
        \begin{equation}
            E=\frac{k}{2} \mathbf{w}B^T B\mathbf{w}^T .
        \end{equation}


         In the main text we consider inducing potential differences on the edges via an external gauge field the vector potential $\mathcal{A}$. This can be achieved in the vector version of the problem via the action of a tensor field. More concretely in mechanics this can be the action of a strain field  $\epsilon$. If we consider populating the edges of a given lattice we can represent this edges as vectors $l_{ij}^\alpha$. If one then populates/realizes that edge then we assign a flow
        \begin{equation}
           \mathbf{w}^\alpha_{ij} = \epsilon_{\alpha\beta} l_{ij}^\beta .           
        \end{equation}
        For example, on a square lattice with edges of length $l$ the assigned flows are $w_{ij}^\alpha= \epsilon_{\alpha,x} l$ for the horizontal bonds and $w_{ij}^\alpha= \epsilon_{\alpha,y} l$ for the vertical bonds.

\section{Numerical simulations}

    \subsection{Monte Carlo simulation of the lattice model}
    We consider a diluted square lattice of size $L$ by $L$ with fluctuating particle occupancy. Each site on the lattice is either occupied or unoccupied ($n_i = 1\ \text{or}\ 0$), and each pair of nearest neighbors $(i,j)$ is connected by a single edge with an assigned quantity $w_{ij}$. The quantity $w_{ij}$ is chosen according to our definition of model A, B, C, which is described in details below. Let $N=\sum_i n_i$ be the the occupation number,\ $N_e$ the number of edges and $\mathbf{w}$ be the $N_e \times 1$ vector containing all the $w_{ij}$. The total energy of the system is then
    \begin{equation}\label{eq:energy}
        E=  -\mu N + \alpha \vert B\mathbf{w} \vert^2,
    \end{equation}
    where $\mu$ is the chemical potential, $\alpha$ is the frustration strength, and $B$ is the circuit matrix. 
    
    We perform Monte Carlo (MC) simulation of our lattice models at fixed chemical potential and fixed temperature. We start with an empty lattice and use open boundary conditions. 
    We use local dynamics to update our system in order to mimic the growth process of GFAs.  
    The local trial moves are as follows: at each MC step, we randomly select a site, and
    \begin{enumerate}[label=(\roman*),leftmargin=0pt,itemindent=1.5em]
        \item if the site is occupied, we remove the particle from the site; also, we remove all the existing edges between the removed site and its occupied nearest neighbors.
    
        \item if the site is unoccupied, we put a particle on it; also, we draw edges between the newly occupied site and its occupied nearest neighbors, and assign $w_{ij}$ to the edges. The assignment protocol for the $w_{ij}$ is different for model A, B, and C:
        \begin{itemize}
            \item \textit{Model A}: every time an edge $(i,j)$ is created, $w_{ij}$ is randomly chosen to be $+1$ or $-1$ with equal probabilities.
            \item \textit{Model B}: $w_{ij}$ is determined from a classical, time-dependent gauge field, whose divergence is zero and gives rise to a uniform magnetic field  perpendicular to the lattice plane with constant time variations; we choose Landau gauge in our simulation:
            \begin{equation}
                \vec{\mathcal{A}}(\mathbf{r},t) = \mathcal{B}_z(t) x \mathbf{\hat{y}}.
            \end{equation}
            The $w_{ij}$ at position $\mathbf{r}=(x,y)$ is obtained via:
            \begin{equation}
                w_{ij} =- \dot{{\mathcal{A}}}_{ij} \, , \, \,\, \dot{{\mathcal{A}}}_{ij} \equiv \dot{\vec{\mathcal{A}}} \cdot \vec{l}_{ij} = 
                \begin{cases}
                        0  & \text{if    } \vec{l}_{ij} \parallel \mathbf{\hat{x}} \\
                        x  & \text{if    } \vec{l}_{ij} \parallel \mathbf{\hat{y}}
                \end{cases},
            \end{equation}
            where $\vec{l}_{ij}$ is the unit vector pointing from node $i$ to node $j$. Note that the time changing rate of $\mathcal{B}_z(t)$ has been absorbed into the frustration strength $\alpha$: $\alpha \sim \mathcal{\dot{B}}^2_z(t)$.
            \item \textit{Model C}: combination of model A and B: every time an edge $(i,j)$ is created,  $w_{ij}$ is chosen to be $-\dot{\mathcal{A}}_{ij}+\gamma$ or $-\dot{\mathcal{A}}_{ij}-\gamma$ with equal probabilities. Where $\gamma$ is the relative strength of random frustration to gauge field.
        \end{itemize}
    \end{enumerate}
    
\begin{figure}
        \centering
        \includegraphics[width=0.4\textwidth]{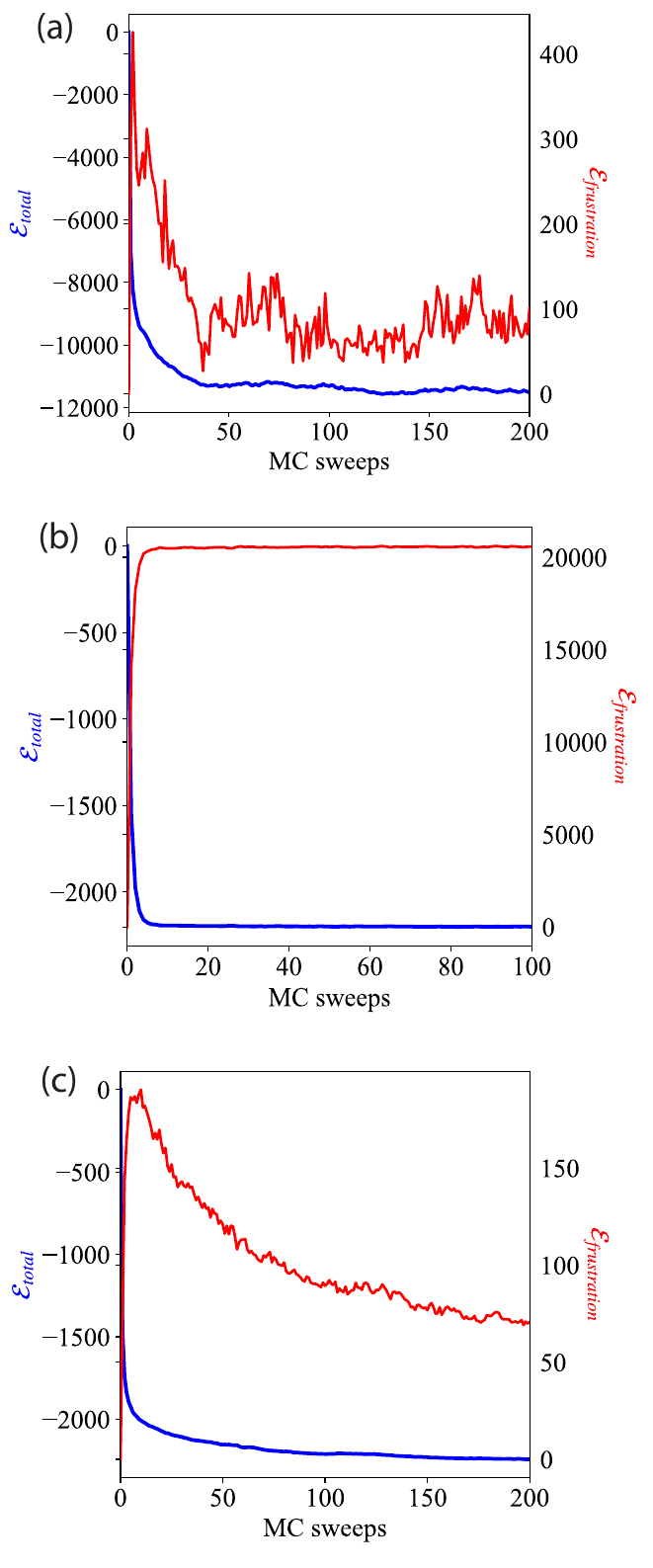}
        \caption{Total energy (blue) and frustration energy (red) as
        function of MC sweeps for (a) Model A, $(\beta\mu, \beta\alpha)=(5,4)$; (b) Model B, $(\beta\mu, \beta\alpha)=(10,0.1)$; (c) Model C, $(\beta\mu, \beta\alpha)=(10,0.1)$, $\gamma = 6$.}
        \label{fig:MCdynamics}
    \end{figure}

    After each trial move, we evaluate the energy change $\Delta E$ based on Eq.~\eqref{eq:energy}, and accept the move with probability $P_{accept} = \min(1, \exp{(-\Delta E / T}))$ according to the Metropolis algorithm. For Model A and C, the simulation is run for 200 MC sweeps to obtain the final configuration, where 1 MC sweep = $L^2$ MC steps; for Model B, the simulation is run for 100 MC sweeps to obtain the final configuration. In Fig.~\ref{fig:MCdynamics} we show examples of energy as function of MC sweeps.

    \subsection{Identifying phase boundaries}

\begin{figure}
        \centering
        \vspace{-0.03\linewidth}
        \includegraphics[width=0.35\textwidth]{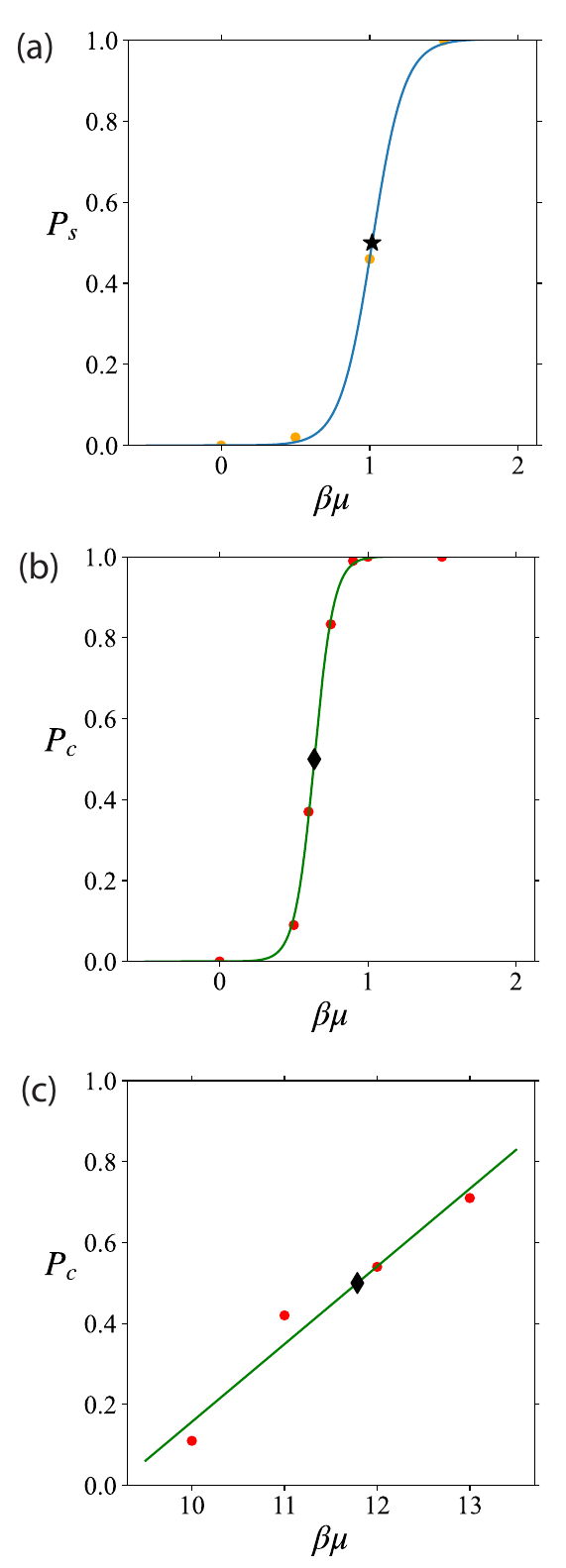}
        \caption{Fitting of structure percolation probability $P_s(\beta\mu, \beta\alpha)$ for (a) Model A. $\beta\alpha=1$. Fitting of current percolation probability $P_c(\beta\mu, \beta\alpha)$ for (b) Model B, $\beta\alpha=0$ and (c) Model B, $\beta\alpha=2$. Linear fit is used when the range of the data points are relatively narrow and cannot be well-fitted by sigmoid function.}
        \label{fig:fitting}
    \end{figure}

    We locate the transition point $\mu_c$ at each $\alpha$ for $L=50$ lattices. For each $(\beta\mu, \beta\alpha)$, we generate 100 independent samples, and take the percentage of percolating samples as the percolation probability $P(\beta\mu, \beta\alpha)$. The criteria for structure (current) percolation is whether a spanning cluster (spanning cluster which carries nonzero current) exists. We use sigmoid or linear function to fit $P(\beta\mu, \beta\alpha)$, and locate $\mu_c$ by $P(\beta\mu_c, \beta\alpha)=0.5$. These transition points are shown in Fig.~2 in the main text and are used to fit the phase boundary. In Fig.~\ref{fig:fitting} we show examples of sigmoid and linear fit for structure and current percolation.

    \subsection{Calculating correlation functions}
    We calculate real space current-current correlation function as $C_{\mu\nu}(r) = \langle i_\mu(0) i_\nu(r) \rangle$. In order to avoid the effect that particles accumulates at boundary, we first cut the lattice by $\Delta L = 10$ (so $L'=30$) and only consider particles in the bulk. The correlation function is first averaged across the sample and then averaged over 200 independently generated samples.

    To calculate current-current correlation functions in Fourier space, we first Fourier transform the current vector:
    \begin{equation}
        i_\mu (\mathbf{q}) = \sum_{e=1}^{N_e} i_{\mu,e} \exp{(j\mathbf{q}\cdot \mathbf{r}_e)},
    \end{equation}
    where $N_e$ is the total number of edges in the system, $j=\sqrt{-1}$, and $\mathbf{r}_e$ is the vector pointing from the center of the lattice to edge $e$. Note that we have ignored the constant factor $1/V$ for simplicity, where $V$ is the total volume of the lattice. The correlation function in Fourier space is then
    \begin{equation}
        C_{\mu\nu}(\mathbf{q}) = \langle i_\mu(-\mathbf{q}) i_\nu(\mathbf{q}) \rangle,
    \end{equation}
    where $C_{\mu\nu}(\mathbf{q})$ is averaged over 200 independently generated samples. We also put a lower $q$ cutoff: $q_{min} = 2\pi / L'$, so that we only consider correlations over length scales shorter than the system size.

    Additionally, we calculate the angular and radial dependence of $C_{\mu\nu}(\mathbf{q}) \sim h(\theta)f(q)$ in polar coordinates $(q,\theta)$. To calculate the angular dependence $h(\theta)$, we choose an arc sector with a small angular bin $[\theta-\Delta\theta, \theta+\Delta\theta)$ , where $\Delta \theta \approx 5^{\circ}$. $h(\theta)$ is then calculated by averaging $C_{\mu\nu}(q,\theta)$ over the arc sector. To calculate the radial dependence, we choose a small slice with $[q-\Delta q,q+\Delta q)$ and $[\theta-\Delta\theta, \theta+\Delta\theta)$, where $\Delta q \approx 0.5a$, and the radial part at this slice is given by $C_{\mu\nu}(q,\theta)/h(\theta)$. $f(q)$ is then calculated by averaging $\theta$ over $[0,2\pi)$.
    
\begin{figure*}[ht]
        \centering
        \includegraphics[width=1.0\textwidth]{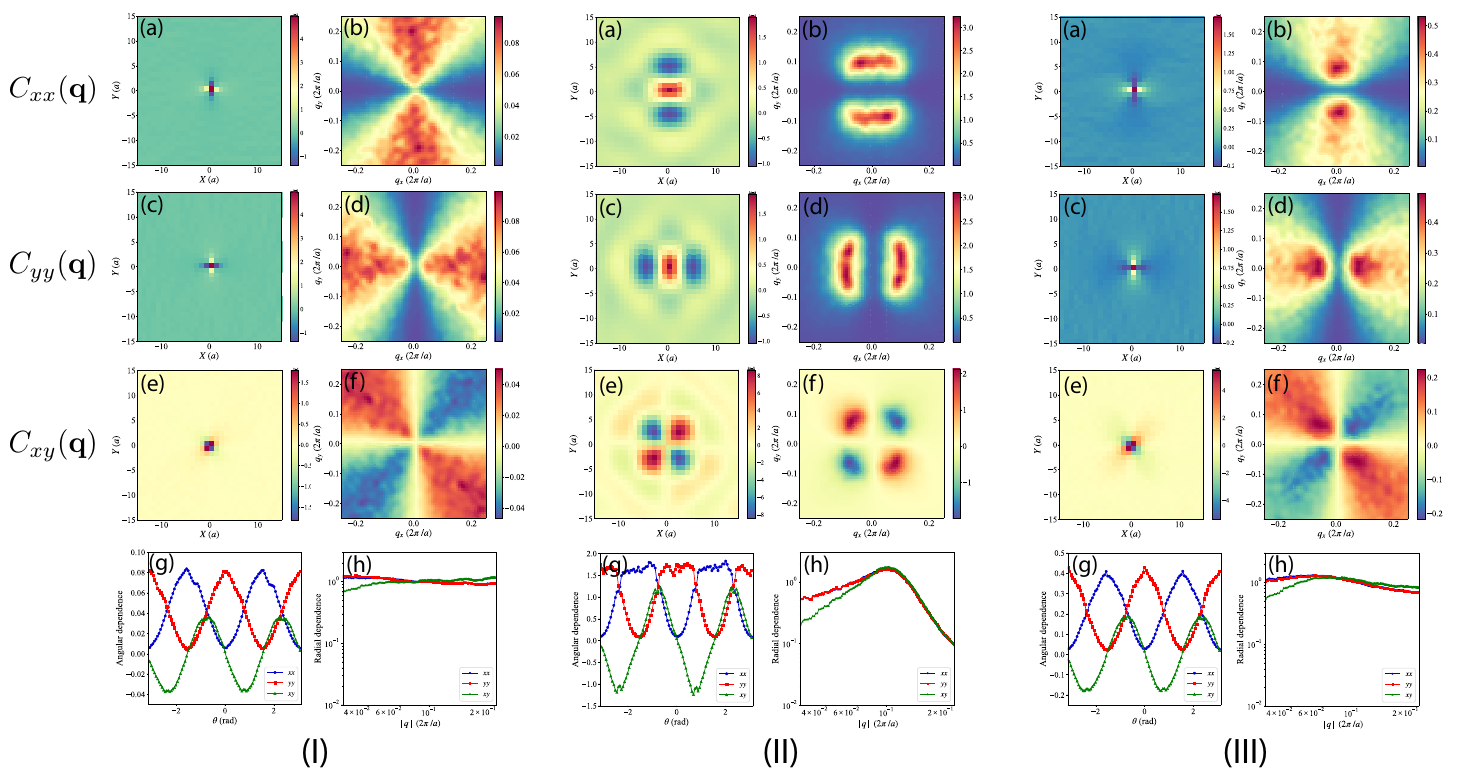}
        \caption{Current-current correlations in real space (left column) and Fourier space (right column) for (I) Model A; (II) Model B; (III) Model C. The parameters are $(\beta\mu, \beta\alpha) = (10,0.1)$, and $\gamma=6$ for Model C. The angular and radial dependence are shown at the bottom row.}
        \label{fig:correlations}
    \end{figure*}

    In Fig.~\ref{fig:correlations} we show current-current correlations $C_{xx}(\mathbf{q})$, $C_{yy}(\mathbf{q})$, and $C_{xy}(\mathbf{q})$ in both real space and Fourier space for all the three models, along with their angular and radial dependence in Fourier space.

    \subsection{Current decomposition of Model C at large $\gamma$}

\begin{figure*}
        
        \includegraphics[width=\textwidth]{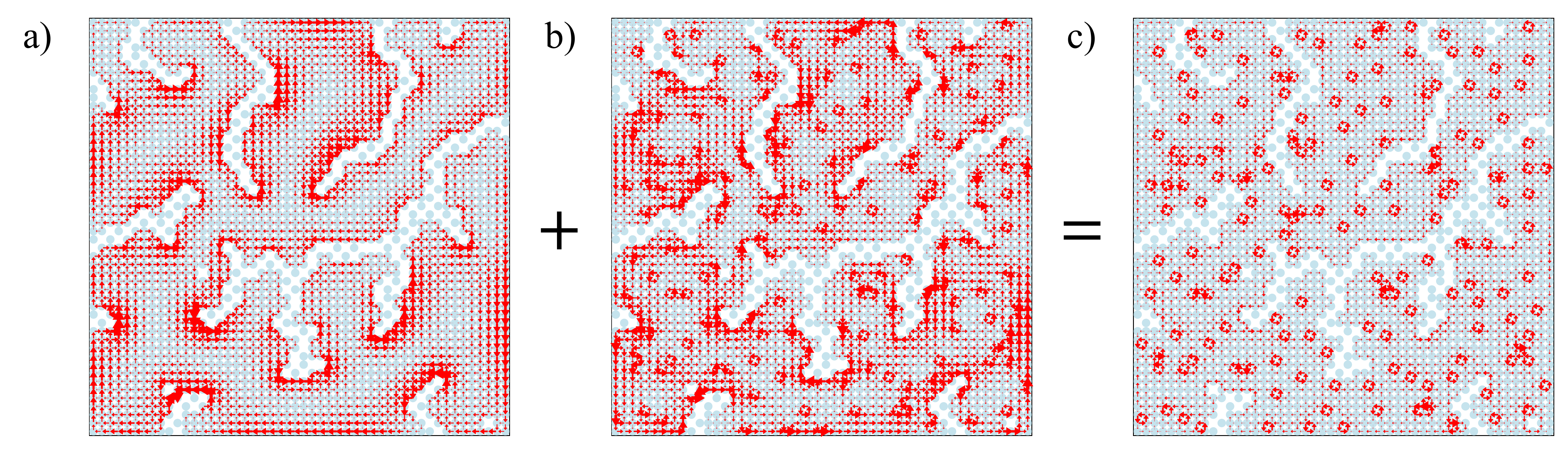}
        \caption{Example of currents from gauge field and random voltage sources cancelling out in Model C at large $\gamma$. a) The current due to the gauge field $\mathbf{w}_{\mathcal{A}}$. b) The current due to the random voltage sources $\mathbf{w}_{\gamma}$. c) The total current which is a sum of the a) and b). Note how the overall magnitude of the current in panel c) is less than the others.}
        \label{fig:ccanceling}
    \end{figure*}

    Here we  show how the disorder introduced by the random voltage sources plays the role of ``defects" that serve to alleviate the stress, in this case current, that necessarily originates from frustration. To do this we take advantage of the fact that we are dealing with a linear problem where the total current can be written as a sum of the two contributions $r\mathbf{i}= B^TB \mathbf{w}_{\mathcal{A}} + B^TB \mathbf{w}_{\gamma}$ where $\mathbf{w}_{\mathcal{A}}=-\dot{\vec{\mathcal{A}}} \cdot \vec{\ell}_e$ and $\mathbf{w}_{\gamma}=\pm \gamma$. In Fig.~\ref{fig:ccanceling} we show how this two contributions in general partially ``cancel out" resulting in a decreased current overall. 


\bibliographystyle{ieeetr}

\providecommand{\noopsort}[1]{}\providecommand{\singleletter}[1]{#1}%

\end{document}